\documentclass[showpacs,aps,prb,twocolumn]{revtex4}
\usepackage{amsmath}
\usepackage{amsfonts}
\usepackage{amssymb}
\usepackage{amsthm}
\usepackage{graphicx}
\usepackage{color}
\begin{document}


\title{Floquet spin states in graphene under ac driven spin-orbit interaction}
\author{A. L\'{o}pez$^{1,2}$}
\email[To whom correspondence should be addressed. Electronic
address: ]{alexander.lopez@physik.uni-regensburg.de}
\author{Z. Z. Sun$^{1,3}$}
\author{J. Schliemann$^1$}
\affiliation{ Institute for Theoretical Physics, University of
Regensburg, D-93040
Regensburg, Germany\\
2. Centro de F\'{i}sica, Instituto Venezolano de Investigaciones
Cient\'{i}ficas, Apartado 21874, Caracas 1020-A, Venezuela\\
3. School of Physical Science and Technology, Soochow University, 
Suzhou, Jiangsu 215006, China
}
\date{\today}

\begin{abstract}
 We study the role of periodically driven time-dependent Rashba 
spin-orbit coupling (RSOC) on a monolayer graphene sample. After recasting the 
originally $4\times 4$ system of dynamical equations as two time-reversal 
related two-level problems, the quasi-energy spectrum and the related dynamics 
are investigated via various techniques and approximations. 
In the static case the system is a gapped at the Dirac point.
The rotating wave approximation (RWA) applied to the driven system 
unphysically preserves this feature, while the Magnus-Floquet approach as 
well as a numerically
exact evaluation of the Floquet equation show that this gap is
dynamically closed.
In addition, a sizable oscillating pattern of the out-of-plane spin 
polarization is found in the driven case 
for states which completely unpolarized in the static limit. 
Evaluation of the autocorrelation function shows that the original uniform 
interference pattern corresponding to time-independent RSOC gets 
distorted. The resulting structure can be qualitatively explained as a 
consequence of the transitions induced by the ac driving among the static 
eigenstates, i.e., these transitions modulate the relative phases that add up 
to give the quantum revivals of the autocorrelation function. 
Contrary to the static case, in the driven scenario, quantum revivals 
(suppresions) are correlated to spin up (down) phases. 
\end{abstract}
\maketitle
\section{introduction}
 
One of the key features of relativistic (massless) free 
particle states is that they evolve, at least in effectively one-dimensional 
situations, in time without spreading.  
This in turn relies on the linear nature of the relativistic dispersion 
relation which for photons reads $\omega(k)=ck$, 
with c the speed of light. The condensed matter relativistic-particle 
analog is found in the low energy approximation 
(long wavelength) of single layer graphene where the chiral massless 
particles move with a speed $v_F\approx c/300$ 
\cite{novoselov1}. This linear spectrum provides graphene with remarkable 
transport properties such as high mobility\cite{geim}, 
Klein tunneling\cite{guinearmp} and unconventional spin Hall 
effect\cite{sarmarmp,kane}. The later stems from the interplay 
between intrinsic  spin-orbit interaction and the coupling extrinsically 
induced by external gate voltages or an appropriate substrate. This extrinsic 
coupling\cite{macdonald,guinea,stauber}, so called Rashba spin-orbit
(RSOC) has been also found to give rise to spin polarization \cite{rashbapol}  
and relaxation \cite{fabianrelax,guinearelax} effects. 

Although the role of static RSOC on graphene has been extensively 
discussed in the literature, to the best of our knowledge, 
the role of periodically driven time dependent RSOC on graphene samples 
has not been analysed so far. Yet, recent works have focused 
on the dynamical features of charge currents induced by means of 
time dependent extrinsic spin-orbit interaction on mesoscopic 
semiconductor quantum rings where Rabi oscillations are shown to appear as 
well as collapse and revival phenomena\cite{peeters1}. The main motivation of our work is twofold: First at all, we are interested in determining the feasibility of ac driven fields to generate and modulate a finite spin polarization of carriers in graphene for states that under static conditions remain unpolarized. In addition, we are also interested in the dynamical modulation of the effective Rashba coupling strength $\Lambda=\lambda/\hbar\Omega$ which would allow to explore regimes beyond the static limit domain.

Taking advantage of the periodicity of the problem, the evolution equations 
can be solved via Floquet theory. A standard approach here consists of 
expressing the Hamiltonian in a Fourier mode expansion leading to an 
infinite-dimensional eigenvalue problem for the so-called quasi-energies 
\cite{milena,chu}. This quasi-energy spectrum carries nontrivial information 
on the topological nature of the system under study\cite{topological1}, 
and for semiconductor quantum wells with a zincblende structure it has been 
recently shown that ac driving can induce a topological phase 
transition\cite{fti}. 

Practically, in order to treat the infinite 
eigenvalue problem one has to truncate at an order of the harmonic 
expansion chosen appropriately to yield well-converged results.
An alternative approach to the Floquet problem which does not rely on
Fourier expansions has been devised by Magnus\cite{magnus1}. 
This method appears to be somewhat less popular and amounts in formulating
the time evolution operator as the exponential of a series of nested 
commutators. It has the virtue of both preserving unitarity at any order in 
the series expansion (in contrast to truncation of the Dyson series within a 
perturbative approach) and avoids the infinite-dimensional 
eigenvalue problems. Following\cite{magnus2} we will make use of the Magnus 
expansion approach combined with Floquet theory in order to 
generate semi-analytical solutions 
of the dynamics induced by periodic RSOC.

Since RSOC couples the spin and pseudospin degrees of freedom the problem
is, for a given wave vector, generically 
four-dimensional. However, by an appropriate unitary transformation,
the evolution equations can be recast as a set of two equivalent two-level 
Schr\"{o}dinger equations related by time reversal. 
In this way we can explicitly analyse which static 
states get dynamically coupled.

Our main results are the following: 
The ac driven RSOC induces a quasi-energy 
spectrum where the original gap due to static spin-orbit coupling is 
dynamically closed. In particular, at the Dirac 
point ($\vec k=0$) the dynamics is exactly solvable with zero quasi-energy 
Floquet states. This quasi-energy spectrum is  two fold-degenerate 
as a consequence of time reversal invariance of spin-orbit interaction, 
and the closing of the original gap is due to  the destructive 
interference induced among the initially uncoupled positive and negative 
energy RSOC eigenstates. Then we show that sizeable alternating out of 
plane spin polarization ensues on states that under static conditions remain 
unpolarized. We also find that the uniform interference 
pattern shown by the autocorrelation function for static RSOC gets distorted 
due to the interlevel mixing of the static eigenstates which dynamically 
modulates the relative phases that add up in the quantum revivals of the 
autocorrelation function. In the driven case quantum revivals (suppressions) 
are directly 
correlated to spin up (down) phases of the out of plane spin polarization.
Since the autocorrelation function is related to the Fourier transform of the 
local density of states\cite{kramer}, and because spin probes can be more 
demanding in practical implementations than charge detection,
its evaluation yields useful indirect 
information on the spin degree of polarization. 
We believe these findings 
have the potential to provide interesting new strategies to dynamically 
control spin 
properties of charge carriers in graphene for future spintronics applications.  

The paper is organized as follows. In section {\rm II} we describe the 
spectum for static RSOC and introduce the model hamiltonian for periodically 
driven RSOC. Here we also present the exact solution to the dynamical 
equations corresponding to the Dirac point $k=0$.  The main results of the 
Floquet-Magnus approach for the semi-analytical solution of the evolution 
operator at finite momentum are presented in section {\rm III}. Next, in 
section {\rm IV} we compare the quasi-energy spectrum obtained through Magnus 
approach to that given by making a rotating wave approximation. We also 
evaluate and discuss the out of plane spin polarization as well as the 
onset of quantum revivals for the autocorrelation function. Finally, in 
section {\rm V} we give some concluding remarks and discuss an experimental scenario where our results could be tested.
\section{model}
We consider a graphene monolayer sample subject to periodic time dependent spin-orbit interation of the Rashba type. In graphene RSOC interaction emerges as a consequence of $\sigma$ and $\pi$ orbital mixing\cite{yao} and stems from the  induced electric field due to the substrate 
over which the graphene sample lies or by applied gate voltages. Then, a periodic modulation can in principle be implemented by means of time dependent gate voltages or by the induced time varying 
electric field within a parallel plates capacitor coupled to a LC circuit. Under these circumstances the  induced RSOC perturbation could  be 
given a periodic time dependence $V(t)=\lambda(t)\vec{s}\centerdot(\hat{z}\times\vec{\sigma})$, where the driving amplitude will be assumed 
to be periodic $\lambda(t+T)=\lambda(t)$ with $\lambda(0)=\lambda_R$, the coupling strenght in the static case. 

Concerning energy scales the value of the intrinsic and extrinsic spin-orbit coupling parameters $\Delta$ and $\lambda_R$ in graphene have been  
obtained  by tight binding \cite{guinea,zarea} and  band structure calculations \cite{macdonald,yao}. They gave estimates in the range  
$10^{-6}-10^{-5} {\rm e V}$, much smaller than any other energy scale  in the problem (kinetic, interaction and disorder). However, the RSOC strength has recently 
been reported\cite{rashbaexp} to be of order $\lambda_R\approx 0.2\tau$,  where  $\tau\approx 2.8 {\rm e V}$ is the value of the first-neighbor hopping 
parameter for graphene within a tight binding approach.

The formulation of the problem is as follows. In momentum space and and taking into account the energy scales of spin-orbit coupling we can work within the so called long wavelength approximation, where the total hamiltonian for monolayer
graphene in presence of time dependent RSOC can be described by the $8\times8$ hamiltonian\cite{kane}
\begin{equation}
 \mathcal{H}(\vec{k},t)=(\sigma_x\tau_zk_x+\sigma_y\tau_y)s_0+\lambda(t)(\sigma_x\tau_zs_y-\sigma_ys_x),\label{dirac0}
\end{equation}

where $v_{F}\sim 10^6 {\rm m/s}$ is the Fermi velocity in graphene, $\vec{\sigma}=(\sigma_x,\sigma_y,\sigma_z)$ is a vector of Pauli 
matrices, with $\sigma_z=\pm1$ describing states on the sublattice $A(B)$ and so called pseudo spin degree of freedom, whereas
$\tau_z=\pm1$ describes the so called Dirac points ${\bf K}$ and ${\bf K}'$, respectively. In addition, $\vec{k}=(k_x,k_y)$ is the momentum 
measured from the ${\bf K}$ point and $s_i$  ({\rm i=0,x,y,z}) represents the real spin degree of freedom, with $s_0$ the identity matrix. In 
addition, $\lambda(t)$ gives the time dependence of the RSOC and we have neglected the intrinsic spin-orbit contributions. Now since RSOC 
does not mix the valleys, we can focus on any of the two Dirac points, say ${\bf K}$  and then the results for the ${\bf K'}$ point are found 
by the substitution $k_x\rightarrow -k_x$. Yet, we will formulate the problem in an isotropic way such that the results for the
${\bf K'}$ Dirac point will inmediately follow. Before dealing with the time dependent problem we summarize the main results for static RSOC. 

The spectrum of the noninteracting  Hamiltonian 
\begin{equation}
\mathcal{H}_0=\hbar v_{F}\vec{\sigma}\centerdot\vec{k} 
\end{equation}
is given by the linear 
dispersion relation 
\begin{equation}
\epsilon^0_{\sigma}(k)=\sigma\hbar v_F\sqrt{k^2_x+k^2_y}\equiv\sigma\hbar v_F k,
\end{equation}
whereas its eigenbasis is 
spanned by the spinors 
\begin{equation}
|\phi_\sigma(\vec{k})\rangle=\frac{1}{\sqrt{2}}\left(
\begin{array}{c}
 1\\
\sigma e^{i\theta}
\end{array}
\right)
\end{equation}
with $\tan\theta=k_y/k_x$, and 
$\sigma= 1$ ($-1$) describes electron (hole) states. 

When a static RSOC interaction term is present the hamiltonian near the 
$\bf{K}$ point reads
\begin{equation}
\mathcal{H}(\vec{k})=\hbar v_F(\sigma_xk_x+\sigma_yk_y)s_0+\lambda_R(\sigma_xs_y-\sigma_ys_x)\label{dirac01}
\end{equation}
and the energy spectrum changes to $\pm\epsilon_\pm$ with
\begin{equation}\label{rso}
\epsilon_{\pm}(k)=\pm\lambda_R+\sqrt{\lambda^2_R+(\hbar v_F k)^2}.
\end{equation}

Since RSOC  mixes the $\sigma$ and $\pi$ atomic orbitals it induces a gap 
$\delta_0=2\lambda$ at the Dirac point $k=0$. 
The static Rashba hamiltonian in Eq.(\ref{dirac01}) is diagonalized by the unitary transformation  $U(\vec{k})$ given explicitly as
\begin{widetext}
 \begin{equation}\label{unitary}
U(\vec{k})=
\frac{1}{\sqrt{2}}\left(
\begin{array}{cccc}
- i e^{i\theta}\sin\gamma_+&-\cos\gamma_+&i\cos\gamma_+& e^{-i\theta}\sin\gamma_+ \\
- i e^{i\theta}\sin\gamma_-&\cos\gamma_-&-i\cos\gamma_-& e^{-i\theta}\sin\gamma_- \\
 i e^{i\theta}\sin\gamma_-&-\cos\gamma_-&-i\cos\gamma_-& e^{-i\theta}\sin\gamma_- \\
 i e^{i\theta}\sin\gamma_+&\cos\gamma_+&i\cos\gamma_+& e^{-i\theta}\sin\gamma_+ 
\end{array}
\right),
\end{equation}
\end{widetext}
where $\cos\gamma_\pm=\epsilon_\pm/\sqrt{(\hbar v_F k)^2+\epsilon^2_\pm}$. In this basis the static RSOC hamiltonian reads
\begin{equation}
\tilde{\mathcal{H}}(\vec{k})=\textrm{Diag}\{-\epsilon_+,\epsilon_-,-\epsilon_-,\epsilon_+\}.
\end{equation}
This particular choice of basis will simplify the calculations that follow. 

We are interested in analyzing the emergent dynamics of Dirac fermions in monolayer graphene when the amplitude of RSOC is  a periodically varying function of time $\lambda(t)=\lambda_R\cos(\Omega t)$, 
with $\lambda_R$ and $\Omega$ the amplitude and frequency of the driving term. Then we would have to deal with the following $4\times4$ evolution equations
\begin{equation}\label{time1}
i\hbar\partial_t\Psi(\vec{k},t)=\mathcal{H}(\vec{k},t)\Psi(\vec{k},t)
\end{equation}
However, if we make use of the unitary transformation (\ref{unitary}) the time dependent hamiltonian (\ref{dirac0})
becomes isotropic and block-diagonal
\begin{equation}\label{interaction}
\tilde{\mathcal{H}}(k,t)=\left(
\begin{array}{cc}
 h_-(k,t)&0 \\
0 & h_+(k,t)
\end{array}
\right),
\end{equation}
with both sub-blocks periodic functions of time i.e. $h_\pm(k,t+T)=h_\pm(k,t)$. Therefore the unitary transformation 
represented by (\ref{unitary}) simplifies considerably the mathematical resolution 
of the evolution equations by recasting the problem as two time reversal pairs of coupled $2\times 2$ two level problems. In addition, 
it has the physical appealing feature of clearly giving the subset of states which are dynamically 
coupled through the time dependent interaction.

Let us then focus on the upper block $h_-(k,t)$ which reads
\begin{widetext}
\begin{equation}\label{interaction0}
h_-(k,t)=-\frac{2}{\epsilon_-+\epsilon_+}\left(
\begin{array}{cc}
 (\hbar v_F k)^2+\lambda(t)\epsilon_+&\hbar v_F k [\lambda_R-\lambda(t)]\\
\hbar v_F k [\lambda_R-\lambda(t)]&-(\hbar v_F k)^2+\lambda(t)\epsilon_-
\end{array}
\right),
\end{equation}
\end{widetext}
whereas the lower block is obtained by changing the sign of the amplitude $\lambda_R\rightarrow-\lambda_R$. Because of this symmetry relation among the two subspaces their quasi-energy spectra are identical. This is to be expected since RSOC is time reversal invariant.

First of all, we notice that in the static limit $\lambda(t)\rightarrow\lambda_R$ the reduced hamiltonian (\ref{interaction}) is 
block diagonal
\begin{equation}
h_-(k)=\left(
\begin{array}{cc}
 -\epsilon_+&0 \\
0&\epsilon_-
\end{array}
\right)
\end{equation}
We also note that at the Dirac point $k=0$, one gets
\begin{equation}
h_-(0,t)=\left(
\begin{array}{cc}
 -2\lambda(t)&0 \\
0&0
\end{array}
\right).
\end{equation}
In this case the resulting dynamics 
\begin{equation}
i\hbar\partial_t|\phi(t)\rangle=h_-(0,t)|\phi(t)\rangle
\end{equation}
is exactly solved by the eigenspinors 
\begin{eqnarray}\label{zeroenergy}
|\phi_1(t)\rangle&=&(e^{2if(t)},0)\\ \nonumber
|\phi_2(t)\rangle&=&(0,1)
\end{eqnarray}
where 
\begin{equation}
f(t)=\frac{1}{\hbar}\int^t_0d t'\lambda(t').
\end{equation}
As will be discussed below, these solutions correspond to zero quasi-energy Floquet states. 
The corresponding evolution operator is diagonal and given as
\begin{equation}\label{exact}
U_-(0,t)=e^{i f(t)}\textrm{Diag}\{e^{i f(t)},e^{-i f(t)}\}.
\end{equation}
\section{Magnus-Floquet approach}
Although we have shown that at $k= 0$ the dynamics is exactly solvable, this is no longer true for finite $k$. Then, we need to resort to approximate solutions. As we discuss below, a semi-analytical approach known as  Magnus-Floquet expansion will be suitable for dealing with the dynamical equations of periodically driven systems.
Since the Magnus-Floquet approach is not so popular in the literature we now briefly summarize its main results (see reference 
\cite{magnus2} for more detailed derivations). 

In the language of differential equations, the matrix solution $S(t)$  of a $n$-dimensional system of dynamical 
evolution equations (here we omit the orbital degrees of freedom for ease of notation), 
\begin{equation}\label{flo1}
\partial_t\Psi(t)=A(t)\Psi(t)
\end{equation}
i.e. a matrix that satisfies
\begin{equation}\label{flo2}
\partial_tS(t)=A(t)S(t)
\end{equation}
is called a fundamental matrix solution if all its columns are linearly independent. If in addition, there is a time $t=t_0$ 
such that $S(t_0)$ is the identity matrix, then $S(t)$ is called a principal fundamental matrix solution. To solve Eq.(\ref{flo2}) Magnus\cite{magnus1} proposed to find exponential solutions to the evolution operator in the form  \begin{equation}
 S(t)=e^{M(t)}
\end{equation}
and then wrote $M(t)$ as an infinite series 
\begin{equation}
M(t)=\sum_{j=1}^\infty M_j(t), 
\end{equation}
where each term $M_j(t)$ is given as a combination of nested commutators, with the first terms reading as
\begin{widetext}
\begin{eqnarray}
M_1(t)&=&\int_0^t A(t_1)dt_1\\
M_2(t)&=&\frac{1}{2}\int_0^tdt_1\int^{t_1}_0[A(t_1),A(t_2)] dt_2\\
M_3(t)&=&\frac{1}{6}\int_0^tdt_1\int^{t_1}dt_2\int^{t_2}_0([A(t_1),[A(t_2),A(t_3)]]+[A(t_3),[A(t_2),A(t_1)]]) dt_3\\
&&\vdots\nonumber
\end{eqnarray}
\end{widetext}
On the other hand, for periodic driving \cite{milena,chu} $A(t+T)=A(t)$, Floquet's theorem states that the principal fundamental solution of the dynamical equations can 
be written as
\begin{equation}
S(t)=P(t)e^{tF}
\end{equation}
where $P$ and $F$ are $n\times n$ matrices, such that $P(t)$ is periodic $P(t+T)=P(t)$ and $F$ is time independent. Floquet's theorem is the time 
dependent analog of Bloch's theorem in solid state physics for spatially periodic structures  and it provides a time 
dependent transformation such that the so called Floquet states evolve according to the time independent matrix $F$. 
This time dependent transformation is implemented by $P(t)$. 

One important remark is in order since although the interaction $A(t)$ is periodic, the corresponding evolution matrix $S(t)$ is 
not, i.e. $S(t+T)\neq S(t)$. In fact $S(T)$ carries indeed non trivial information on the dynamics of the periodic system.  The eigenvalues 
of $F$ are called Floquet exponents $\rho$. These Floquet exponents can be found by diagonalizing $S(T)=e^{TF}$. Yet, they are not 
uniquely defined since $\rho\rightarrow\rho+2in\pi/T$ leaves $S(T)$ invariant.

In order to determine those exponents one standard approach consists of performing an expansion in the (infinite) eigenbasis of 
time periodic functions $\xi_N(t)=e^{iN\Omega t}$ (Fourier modes). In that periodic basis the evolution operator is diagonalized 
and the Floquet exponents $q_n$ are the logarithms of the eigenvalues of the evolution operator evaluated at $t=T$, i.e. $S(T)$. Then, in order to deal with the infinite
eigenvalue problem one resorts to a truncation procedure in order to determine the Floquet exponents. 

The Magnus approach avoids the need to solve the infinite dimensional eigenvalue problem and has the physical virtue of preserving unitarity of the evolved state to any order in the expansion. The connection between Magnus expansion and Floquet's theorem is found in reference \cite{magnus2} where the authors present a solution of 
the evolution equations that consists of writing the periodic part $P(t)$ as an exponential 
\begin{equation}
P(t)=e^{\Omega(t)}\qquad \Omega(t+T)=\Omega(t),
\end{equation} 
and then they proceed to expand both operators $\Omega(t)$ and $F$ in power series
\begin{equation}\label{terms}
\Omega(t)=\sum_{j=1}^\infty \Omega_j(t),\qquad
F=\sum_{j=1}^\infty F_j.  
\end{equation}
Now, since $S(t)$ is by construction a principal fundamental matrix  solution $P(T)=P(0)$ is the identity matrix and one gets for all values of $j$ 
\begin{equation}
F_j=M_j(T)/T.
\end{equation}
Introducing the Bernoulli numbers $B_l$ ($B_0=1$, $B_1=-1/2$, $B_2=1/6,B_4=-1/30\dots$) such that $B_{2m+1}=0$ for $m\ge 1$, the exponent operator term contributions $\Omega_j(t)$ satisfy a recurrence relation in terms of two auxiliary time dependent operators $W(t)$ and $T(t)$, according to the relations
\begin{equation}
 \partial_t\Omega_j(t)=\sum^{j-1}_{l=0}\frac{B_l}{l!}(W^{(l)}_j(t)+(-1)^{l+1}T^{(l)}_j(t))\qquad (n\ge 1),
\end{equation}
In turn, the $W's$ and $T's$ are given through the iterative relations
\begin{eqnarray}
W^{(l)}_j&=&\sum^{j-l}_{m=1}[\Omega_m,W^{(l-1)}_{j-m}]\quad (1\le l\le j-1)\\
T^{(l)}_j&=&\sum^{j-l}_{m=1}[\Omega_m,T^{(l-1)}_{j-m}]\quad (1\le l\le j-1)\\
W^{(0)}_1&=&A,\qquad W^{(0)}_j=0\quad (j>1)\\
T^{(0)}_j&=& F_j,\qquad (j>0).
\end{eqnarray}
In practical calculations, the relations in the last two lines serve to initialize the iterative procedure.
\section{Results and discussion}
In order to apply the previous results to our problem one just has to make the  following identifications 
$S(t)\rightarrow U_-(k,t)$, $A(t)\rightarrow-ih_-(k,t)/\hbar$. Then, the characteristic exponents are proportional to the 
quasi-energies $\rho=-iq_n$.
\begin{figure}[ht]
\begin{center}
\includegraphics[width=9cm]{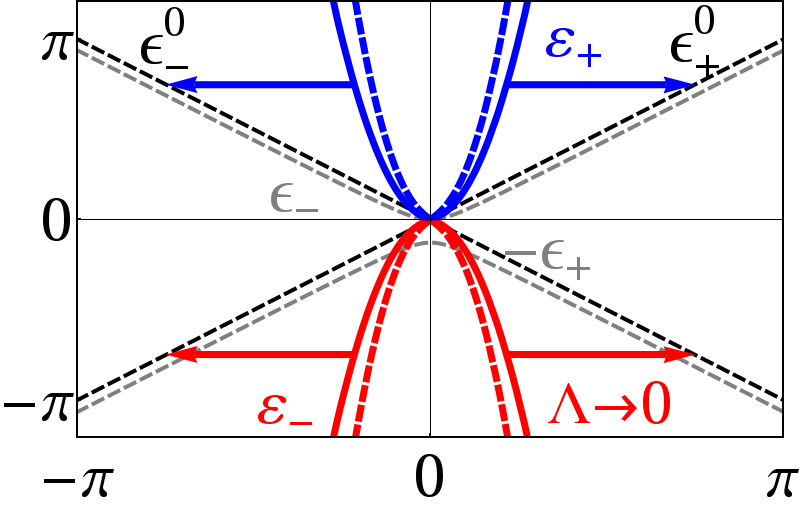} \label{figure1}
\caption{(Color online) First Brillouin zone for the quasi-energy spectrum as function of adimensional non interacting quasi 
particle energy. Due to the dynamical interlevel mixing the static  gap $\delta_0$ gets closed (colored, thick continous lines) as compared  to the static interacting spectrum (gray, thin, dashed lines). Colored arrows depict the limit $\Lambda\rightarrow 0$ where the highly oscillatory contributions tend to cancel and the non interacting spectrum is recovered (black, thin, dashed lines). The Floquet Fourier solutions (colored, thick, dashed lines) show qualitative agreement with the Magnus result however they converge slower for small $\Lambda$. We have expressed all quantities in units of $\tau=2.8{\rm eV}$, the first neighbor hopping parameter within a tight-binding approach and set $\Omega=1$.}
\end{center}
\end{figure} 
For our calculation we choose a time dependence of the form $\lambda(t)=\lambda_R\cos{\Omega t}$, where
 $\Omega$ is the frequency of the driving and $\lambda_R$ the RSOC strength and proceeded to evaluate $F_j$ by the iteration procedure described 
in the previous section. 

We have found a characteristic behavior of the (adimensional) quasi-energies $\varepsilon\pm=q_\pm/\hbar\Omega$,  which qualitatively do not change when one goes beyond third order in the Magnus-Floquet expansion (in the appendix we briefly describe the 
calculations up to fifth order). For finite $\Omega$ they explicitly read as
\begin{eqnarray}\label{quasi}
\varepsilon_\pm&=&\pm\sqrt{\kappa^2 (16\kappa^2\Lambda^2+\Lambda^4-2\Lambda^2 + 1)},
\end{eqnarray}
where we have defined the adimensional quantities $\kappa=v_F k/\Omega$ and $\Lambda=\lambda_R/\hbar\Omega$.

In {\rm FIG.1} we show the typical behavior of the quasi-energies $\varepsilon_\pm$ (blue and red, thick continous lines) given in equations 
(\ref{quasi}) as function of the adimensional quasiparticles non interacting energy $\epsilon^0$. We also show (gray, thin, dashed lines)  
the corresponding static eigenvalues $\epsilon_\pm$ of the RSOC hamiltonian as described by $h_-(k)$, as well as 
 the eigenenergies  of the non interacting hamiltonian (thin, dashed, black lines). We have also included (red and blue, thick, dashed lines) the result from a $20$-mode Fourier expansion of the quasi-energies. We have expressed all quantities in units of the first neighbor hopping parameter $\tau=2.8{\rm eV}$ and for finite $\Omega$ have set $\Omega=1$ and change the effective coupling $\Lambda$. This is true for all the figures within the dynamical case.

 Since  $h_-(k,t)$ mixes the static eigenstates of $h_-(k)$ the relative phases among them are dynamically modulated giving rise 
to interference phenomena and we find that this leads to a dynamical closing of the original gap $\delta_0$. Therefore, the exact solutions at $k=0$ correspond to vanishing
quasi-energies (modulo $\hbar\Omega$). We further find (see arrows in {\rm FIG.1}) that in the limit $\Lambda\rightarrow 0$, corresponding to a highly oscillating field, $\varepsilon_\pm\rightarrow\epsilon^0_\pm$ because then the influence of the driving quickly tends to vanish on average
and intuitively one expects to recover the  non interacting linear spectrum.  The result from a $20-$mode Fourier expansion shows qualitative agreement to the Magnus-Floquet approach,
however a small discrepancy is found and the Magnus result converges faster in the limit of highly oscillating fields $\Lambda\rightarrow 0$, as depicted by thick arrows in {\rm FIG.1}. 

 Within the Magnus-Floquet approach the evolution operator $U_-(\kappa,t)$ corresponding to the hamiltonian $h_-(\kappa,t)$ is found to be given as
\begin{equation}
U_-(\kappa,t)=e^{if(t)}e^{i\vec{V}(\kappa,t)\cdot\vec{p}}e^{i\vec{v}(\kappa,t)\cdot\vec{p}}
\end{equation}
where $\vec{p}$ is a vector of Pauli matrices and the non vanishing components of the vector $\vec{V}$  are given as
\begin{widetext}
\begin{eqnarray}
V_x(\kappa,t)&=&\frac{-\kappa\Lambda\sin{\Omega t}}{{\sqrt{\kappa^2+\Lambda^2}}}(4\kappa^2-2\Lambda^2+1+\Lambda^2\cos{\Omega t})\\
V_y(\kappa,t)&=& 2\kappa\Lambda(1-\cos{\Omega t})\\
V_z(\kappa,t)&=&\frac{\Lambda^2\sin{\Omega t}}{{\sqrt{\kappa^2+\Lambda^2}}}\left(6\kappa^2+1-\kappa^2\cos{\Omega t}\right)
\end{eqnarray}
\end{widetext}
whereas those of $\vec{v}$ are
\begin{eqnarray}
v_x(\kappa,t)&=&\frac{t\kappa\Lambda}{{\sqrt{\kappa^2+\Lambda^2}}}\left(1+4\kappa^2-\Lambda^2 \right)\\
v_z(\kappa,t)&=&\frac{t\kappa^2}{{\sqrt{\kappa^2+\Lambda^2}}}\left(1-5\Lambda^2\right).
\end{eqnarray}
At the Dirac point $\kappa= 0$ we find that $v_x,v_z,V_x,V_y$ all vanish, whereas $V_z=\Lambda\sin{\Omega t}=f(t)$, and thus we recover the exact 
solution (\ref{exact}).

Before we proceed to evaluate other physical quantities of interest we make a brief disgression on the importance of taking into account the full time dependence of the RSOC. In particular, due to smallness of the RSOC strength one can try a rotating wave approximation (RWA) where for a given finite value of $k$ only near to resonance ($\Omega\sim 2 v_F k$) terms are kept in the interacting hamiltonian. Now we show that for the present model this approach gives unphysical results, for instance, it predicts a gap opening at $k=0$. This in turn would imply finite quasi-energy mode contributions at the Dirac point which contradicts the previously described exact result.

Since the results are easily found within the $4$-dimensional formulation we briefly return to the original basis and make a full dimensional discussion. Within a RWA approach, the hamiltonian reads (here we omit the spatial degrees of freedom for ease of notation
\begin{equation}
 \mathcal{H}^{\textrm{RWA}}(t)=\mathcal{H}_0+i\lambda(\sigma_+s_-e^{i\Omega t}-\sigma_-s_+e^{-i\Omega t}),
\end{equation}
which for a given value of $k$ describes near resonance $\Omega\sim 2 v_F k$ spin and pseudo spin flipping processes and neglects the so called secular or counter-rotating terms that oscillate rapidly. In this case the solution is exact and the adimensional quasi-energy spectrum reads
\begin{eqnarray}
\varepsilon^{\textrm{RWA}}_{\sigma s}=\frac{s}{2}\sqrt{\delta^2_{res}+g_\sigma(\kappa)}
\end{eqnarray}
where $\delta_{res}=2\kappa- 1$ describes the resonance and $g_\sigma(\kappa)=\Lambda^2-\kappa+2\sigma\sqrt{\kappa(\Lambda^2+1)+\Lambda^4}$. 
\begin{figure}[ht]
\begin{center}
\includegraphics[width=9cm]{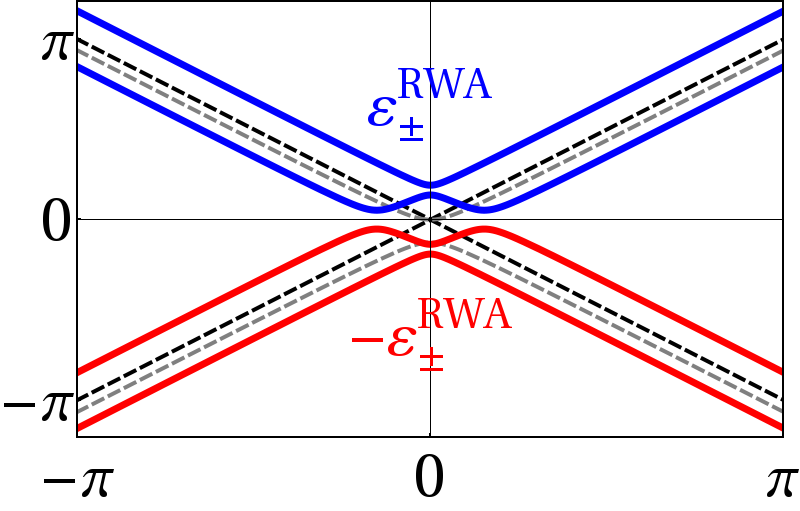} \label{figure2}
\caption{(Color online) Quasi-energies as function of momenta. Gray continuos lines correspond to the static limit. The solution within this approach describes an unphysical gap opened at $\kappa=0$ (see main text).}
\end{center}
\end{figure}
When we evaluate at $\kappa=0$ and finite $\Omega$ one would get the gaps $\Delta=\sqrt{(\Lambda^2+1/4}$ for the originally gapped states and a dynamically opened gap on the static degenerate states $\Delta_{dyn}=1$. These results are in disagreement with the exact solution to the full equations (\ref{zeroenergy}) where we had found zero energy solutions at $\kappa=0$ so this approach is not suitable to describe the dynamical features of the periodic driving. In {\rm FIG.2} we depict the quasi-energy spectrum as a function of non- interacting energies $\epsilon^0$ for this RWA solution.

After this brief discussion we turn back our attention to the Magnus-Floquet solution in order to get additional information on the dynamical behavior induced on the system. This can be found by evaluating some other physical quantities of interest from an experimental point of view. First we analyse the out of plane spin polarization $\mathcal{S}_z(\vec{k},t)=\langle\Psi(\vec{k},t)|S_z|\Psi(\vec{k},t)\rangle$, where $S_z=\hbar/2\sigma_0\otimes s_z$ and $\sigma_0$ the identity matrix in pseudo spin space. Using again the transformation (\ref{unitary}) the out of plane spin polarization reads $S_z(\vec{k})=U(k)S_zU^\dagger(\vec{k})$ and  is found to be isotropic and block anti-diagonal
\begin{equation}
S_z(\vec{k})=\left(\begin{array}{cc}
0&s_-\\
s_+&0
\end{array}
\right)
\end{equation}
with $s_\pm$ given explicitly as
\begin{equation}
s_\pm=-\frac{\hbar}{2\sqrt{\kappa^2+\Lambda^2}}(\kappa p_0\pm i\Lambda p_y)
\end{equation}
with $p_i$ a vector of Pauli matrices and $p_0$ the two dimensional identity matrix. The anti-diagonal structure of $\tilde{S}_z$ in this basis reflects the fact that spin polarization is not conserved in presence of RSOC. 
\begin{figure}[ht]
\begin{center}
\includegraphics[width=9cm]{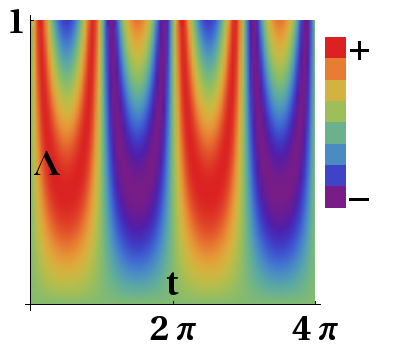} \label{figure3}
\caption{(Color online) Density plot showing the out of plane spin polarization plotted against normalized time $t=\hbar/\tau$ for the exact solution at $\kappa=0$. The periodicity is inherited from the driving field. The vertical axis indicates the normalized adimensional field strength $\Lambda$ and one gets up ($+$) down ($-$) spin components phases equally separated by the zeroes of $f(t)$.}
\end{center}
\end{figure}
\begin{figure}[ht]
\begin{center}
\includegraphics[width=9cm]{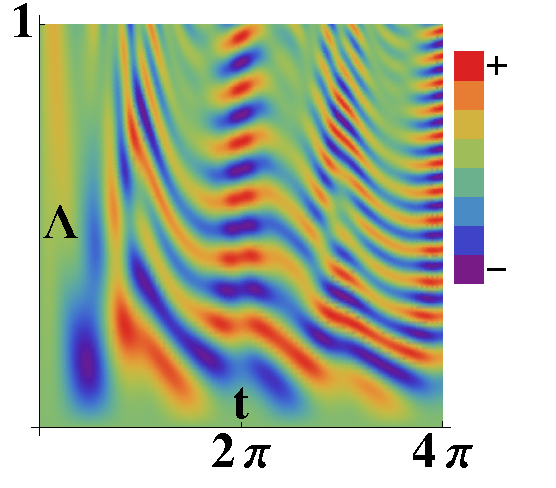} \label{figure4}
\caption{(Color online) Density plot showing the time behavior of the out of plane spin polarization for the semi- analytical Magnus-Floquet solution at finite  $\kappa=1$. Alternating maxima and minima appear due to interlevel mixing among different static eigenstates. This is also manifested by the lost of symmetry with respect to $t=\pi$ and stems from the driven modulated phases since dynamical interlevel mixing leads to interference effects.}
\end{center}
\end{figure}
Then we have to evaluate 
\begin{eqnarray}
\mathcal{S}_z(\vec{k},t)=\langle\Psi(k,0)|U^\dagger(k,t)S_z(k)U(k,t)|\Psi(k,0)\rangle,
\end{eqnarray}
for any initially prepared state $|\Psi(k,0)\rangle$. Next we separate explicitly the four spinor $|\Psi(k,0)\rangle$ in upper and lower components as
\begin{equation}
|\Psi(k,0)\rangle=\frac{1}{\sqrt{2}}\left(\begin{array}{c}
\psi_-\\
\psi_+
\end{array}
\right)
\end{equation}
where $\psi_\pm$ are normalized two dimensional spinors.  After some algebra one gets for the spin polarization in terms of adimensional parameters
\begin{equation}
\mathcal{S}_z(\kappa,t)=\Re\psi^*_-e^{-2if(t)}e^{-i\vec{v}\cdot\vec{p}}e^{-i\vec{V}\cdot\vec{p}}s_-e^{i\vec{V}\cdot\vec{p}}e^{i\vec{v}\cdot\vec{p}}\psi_+/2
\end{equation} 
\begin{figure}[ht]
\begin{center}
\includegraphics[width=9cm]{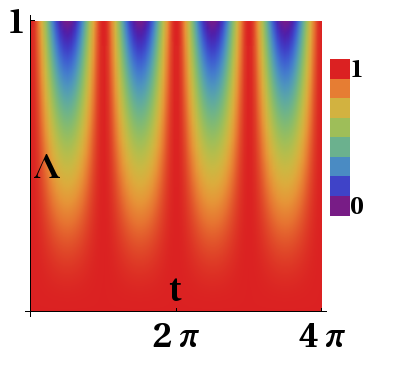} \label{figure5}
\caption{(Color online) Density plot showing the behavior of the absolute value of the autocorrelation function plotted against normalized time $t=\hbar/\tau$ and strength $\Lambda$ for the exact solution at $\kappa=0$. As expected, only for large values of the interaction strength the evolved state departs considerably from the initial state configuration (see main text). }
\end{center}
\end{figure}
\begin{figure}[ht]
\begin{center}
\includegraphics[width=9cm]{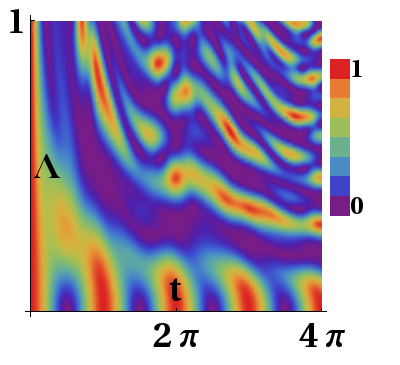} \label{figure6}
\caption{(Color online) Density plot showing the behavior of the absolute value of the autocorrelation function against normalized time $t=\hbar/\tau$ and RSOC strength $\Lambda$ for the Magnus-Floquet solution at $\kappa=1$. The phase interference effects previously discussed. Now the onset of recurrences at small $\Lambda$ is a consequence of interlevel mixing.}
\end{center}
\end{figure}
 For a finite value of $\kappa$ we now choose the initial spinor configuration $\psi_\pm=(\pm i,1)/\sqrt{2}$,  in such way that the out of plane polarization vanishes $\langle|\Psi(k,t)|S_z(k)|\Psi(k,t)\rangle=0$ for a static RSOC. 
 
 In figure {\rm FIG.3} we show a density plot  of the resulting out of plane spin polarization for the exact solution $\kappa=0$. In this case, the only relevant parameter is the adimensional amplitude of the driving field $\Lambda$. As expected, for $\Lambda=0 $ the system remains unpolarized for all values of the adimensional time $t$  (given in units of $\tau/\hbar$). For finite values of the effective coupling an  alternating pattern of spin phases (denoted as $+$ and $-$ representing up and down, respectively) are seen to apper as time evolves. They are symmetrically distributed  among the values $t=n\pi$ where $f(t)$ and thus the relative phases among the static RSOC eigenstates vanish. 
 
 However, as shown in figure {\rm FIG.4}, once $\kappa$ is finite this panorama qualitatively changes. In this case, the additional interference due to level mixing  induces a pattern of alternating maxima $(+)$ and minima $(-)$ for $t_n=n\pi$, $n = \in\mathbb{N}$. The reason for such a behavior is that for a given $t_n$ the evolution operator is given by $e^{TF}$ and thus the polarization maxima and minima $\mathcal{S}(\kappa,T)=\pm1/2$ depend essentially on the quasienergy spectrum properties. Then, increasing $\Lambda$ makes these alternating maxima and minima to get closer. When we move to the next period (corresponding to $t=4\pi$ in figure {\rm FIG.4}) the number of alternating maxima and minima is doubled, and so on. 
 Therefore, dynamical coupling produces a non vanishing out of plane spin polarization with an oscillating time pattern resulting from the mixing of the static eigenstates such that the changing relative phases, modulated by the time dependent interaction prevent total destructive interference to happen. Therefore ac driven RSOC provides a suitable means to dynamically control the degree of spin polarization and could in principle serve to generate non vanishing and non trivial spin polarized phases in otherwise unpolarized states in monolayer graphene.
 
In order to complement the just described physical picture of the spin polarization scenario  we evaluated the autocorrelation function $A(\vec{\kappa},t)$. This is given by the projection of the evolved state $|\Psi(\vec{\kappa},t)\rangle$ along a given (in principle arbitrary) initial spinor configuration $|\Psi(\vec{\kappa},0)\rangle$, i.e. 
\begin{equation}
A(\vec{\kappa},t) =\langle\Psi(\vec{\kappa},0)|\Psi(\vec{\kappa},t)\rangle.
\end{equation}
The absolute value of the  autocorrelation function provides information on the so called recurrences or quantum revivals of the dynamics, i.e. those values of the time parameter for which such overlapping is a maximum. In addition, its Fourier transform is proportional to the local density of states\cite{kramer}.  

In figure {\rm FIG.5} ({\rm FIG.6}) we plot the absolute value of $A(\kappa,t)$ obtained by means of the exact (semi-analytical) evolution operators. For the exact solution shown in figure {\rm FIG.5} only large values of $\Lambda$ induce an appreciable phase change and the system remains mostly correlated to the initial state. As for the case of the out of plane spin polarization, for $t=t_n$ the autocorrelation gives maxima corresponding to red (dark gray) zones in the figure. These signal the return of the system to initial vanishing spin polarization. Maxima and minima of spin polarization correspond to partial quantum revivals. Given the definition of the autocorrelation as the probabibility that the system returns to its initial state  we find that for $\Lambda=0$ giving $\varepsilon=\kappa$ the loci of $A(\kappa,t_n)=1$ correspond to vanishing of spin polarization and are thus equally spaced as $\delta t=\pi$. On the contrary, the maxima and minima of spin polarization correspond in this case to quantum suppressions and are shown as blue (dark) zones in the figure.
As soon as we move towards finite values of $\kappa$ ({\rm FIG.6}) interference phenomena come again into play and we find recurrences represented as read (dark gray) zones  and suppressions, described by purple (black) zones. These arise because of the constructive and destructive interference effects, described previously and are modulated by the ac driving.   
\begin{figure}[ht]
\begin{center}
\includegraphics[width=9cm]{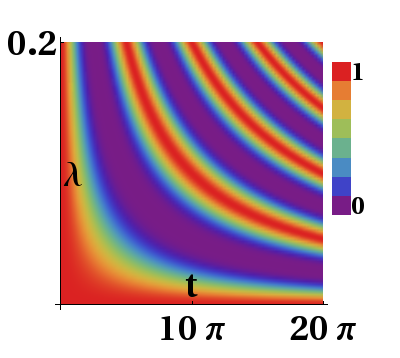} \label{figure7}
\caption{(Color online) Density plot showing the behavior of the absolute value of the autocorrelation function versus normalized time and RSOC strength $\Lambda$ for the static interaction at $\kappa=0$. In this case, quantum revivals are separated by the inverse of the gap $\delta_0=2\lambda$ in accordance to Heisenberg uncertainty principle.}
\end{center}
\end{figure}
\begin{figure}[ht]
\begin{center}
\includegraphics[width=9cm]{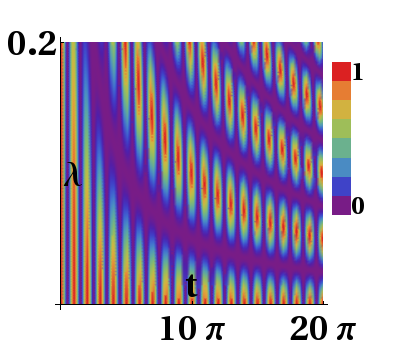} \label{figure8}
\caption{(Color online) For finite values of momentum $\kappa=1$ recurrences are modulated signaling the coexistence of revivals for times inversely proportional to the static gap $\delta_0=2\lambda$ and those corresponding to the larger energy separation $\delta=\sqrt{k^2+\lambda^2}$, which correspond to a smaller time and seen in the figure as modulated alternating pattern of maxima and minima for each recurrence curve corresponding to $k=0$. }
\end{center}
\end{figure}
To check that this physics is different from the static scenario, in Figures {\rm FIG.7} and {\rm FIG.8} we depict the corresponding countour plots for time independent RSOC interaction. For the $\kappa=0$ case shown in {\rm FIG.7}, the alternating pattern of quantum revivals and suppressions is separated by the inverse of the energy gap $\delta_0$. This is due to the fact that the in this situation the relative phase among the non interacting eigenstates is set by the energy separation among them, i.e. $\delta_0$, and according to Heisenberg's time-energy uncertainty relation one should have $\delta t$ inversely proportional to the energy separation. As soon as $\kappa$ is finite, a quasi homogeneous pattern of recurrences is seen. Therefore although in the static case where the spin polarization vanishes for all values of RSOC strength, there appear recurrences indicating there is no longer correlation between $S_z$ and $A(t)$, which for the system under study are only correlated for time dependent driving.

Two comments are in order here. The first one is concerning the validity of the Dirac fermion hamiltonian approximation in order to deal with spin-orbit related phenomena in graphene monolayer. As it is discussed in reference\cite{yao} the large gap of the hamiltonian describing the $\sigma$ orbitals implies that the effective hamiltonian including SOC is essentially given by the long wavelength or Dirac Hamiltonian considered in our model. In addition, M. Gmitra, et al\cite{mitra} used first principles calculations to discuss the relevance of spin-orbit related physics in graphene near and beyond the Dirac points. The second point to remark is the role of localized impurities. As has been discussed in\cite{castro}, the presence of local impurities can enhance the value of the static 
intrinsic spin-orbit coupling strength because of the induced $sp^3$ distortion which leads to a hybridization of the $\pi$ and $\sigma$ orbitals and as shown in reference\cite{mitra} RSOC only respond to this $\sigma-\pi$ hybridization. Therefore, we would expect our results to be robust or even enhanced if localized impurities were included in the model.

Now we would like to compare to other proposals of dynamical modulation of energy gaps under ac drivenin graphene. For instance, Oka and Aoki\cite{oka} found that circularly polarized intense laser fields can induce a photovoltaic effect in graphene, that is, a Hall effect without magnetic fields. This in turn relies on the gap opening at the Dirac point $k=0$. However, as shown recently by Zhou and Wu\cite{wu} in analyzing the optical response of graphene under intense THz fields, the contributions from large momenta make the effective gaps opened to get also dynamically closed. However, it is found that the quasi energy spectrum for the linear polarization leads to linear quasi energy spectrum (see also \cite{foa}). Our model could in principle be mapped into their scenario of linearly polarized radiation field. Yet, in both works \cite{wu,foa} the case of linear polarization leads to linear quasienergy spectrum. Therefore, from the semi parabolic quasi energy spectrum shown in figure (FIG.\ref{figure1}) we can infer that the bending of the quasienergy spectrum makes the spin-orbit driven scenario qualitatively different from these other approaches to dynamical control of graphene electronic properties. Our argument relies on the fact that the topological properties of periodically driven systems are characterized by the quasienergy spectrum\cite{topological1} so we can conclude that the  physics related to ac driven spin-orbit does reveal new physical interesting electronics properties which are absent in the static regime.

\section{conclusions}
We have investigated the role of periodically driven RSOC in monolayer graphene and recasted the original $4$-dimensional problem as an equivalent set of two two-level problems. Due to the induced modulation of the relative phases among the static eigenstates we found a closing of the static gap at the Dirac point $k=0$. This result is in agreement with the available exact solution and differs from RWA where an unphysical gap is seen to appear. This physical picture is confirmed through a Fourier mode expansion and we found that Magnus Floquet approach indeed has the advantage of providing the quasienergy spectrum with less computational effort. We also found that the generation and manipulation of out of plane spin polarization for otherwise spin unpolarized states requires the time driving to be realizable within this set up. Due to the induced interlevel mixing among the static eigenstates we found a set of alternating positive and negative spin phases in clear distinction to the well separated spin phases at the Dirac point corresponding to the exact solution. The dynamical onset of quantum revivals described through the autocorrelation function is directly correlated to the appearance of maxima for either spin phases. However, in the static case, such a correlation does not ensues since the spin polarization vanishes identically whereas the quantum revivals are still present. Concerning the actual experimental realization of our proposal we believe it could be implemented by means of Magnetic Resonance Force Microscopy as reported in\cite{rugal}. Within this scheme, single spin polarization could be detected by means of the frequency shift induced on a cantilever that is used to scan the sample. The sign of the cantilever's frequency shift can be associated to the spin polarization. This detection is achieved by means of low intensity magnetic fields under resonant conditions, thus no magnetic coupling terms need to be included in the description of the dynamics of the charge carriers in graphene. In this way, although in principle RSOC in graphene has a small static strength, the time dependent efective phenomena produce some interesting spin controlling strategies which we believe could provide a route to new implementations of graphene in spintronic devices with appropiate spin detection techniques.

{\it{Acknowledgments}--} 
Z.Z.S. thanks the Alexander von
Humboldt Foundation (Germany) for a grant, and A.L. thanks 
DAAD-FUNDAYACUCHO for financial support.
This work has been supported by Deutsche Forschungsgemeinschaft via GRK 1570.

\appendix
\section{derivation of iterative terms and quasi energy spectrum}
 Using the simplifying 
notation for the time dependent interaction $h^-(k,t_i)=h^-_i$ one gets to fourth order  in the iteration procedure that
\begin{widetext}
\begin{eqnarray}
F_1&=&\frac{-i}{T\hbar}\int_0^T h^-_1dt_1\\
F_2&=&\frac{-i^2}{2T\hbar^2}\int_0^Tdt_1\int^{t_1}_0[h^-_1,h^-_2] dt_2\\
F_3&=&\frac{-i^3}{6T\hbar^3}\int_0^Tdt_1\int_0^{t_1}dt_2\int^{t_2}_0([h^-_1,[h^-_2,h^-_3]]+[h^-_3,[h^-_2,h^-_1]]) dt_3\\
F_4&=&\frac{-i^4}{12T\hbar^4}\int_0^Tdt_1\int_0^{t_1}dt_2\int^{t_2}_0dt_3\int^{t_3}_0([[[h^-_1,h^-_2],h^-_3],h^-_4]+\\\nonumber
&&[[[h^-_3,h^-_2],h^-_4],h^-_1]+[[[h^-_3,h^-_4],h^-_2],h^-_1]+[[[h^-_4,h^-_1],h^-_3],h^-_2]) dt_4\\
\vdots
\end{eqnarray}
\end{widetext}
and performing the corresponding calculations one gets
\begin{eqnarray}
F_1&=&\frac{i\kappa\Lambda}{\sqrt{\kappa^2 + \Lambda^2}}\left(
 \kappa\sigma_z+\Lambda\sigma_x
\right)\\
F_3&=&\frac{-i\kappa\Lambda}{\sqrt{\kappa^2 + \Lambda^2}}\left[
 5\kappa\Lambda\sigma_z-(4\kappa^4-\Lambda^2)\sigma_x
\right]\\
F_5&=&\frac{i\kappa\Lambda}{36\sqrt{\kappa^2 + \Lambda^2}}[
7\kappa\Lambda(-144\kappa^2+31\Lambda^2)\sigma_z\\\nonumber
&&+(576\kappa^4-640\kappa^2\Lambda^2+9\Lambda^4)\sigma_x]
\end{eqnarray}
whereas the even contributions $F_{2j}$ all vanish. The quasienergies $\varepsilon_\pm$ are obtained through diagonalization of the selfadjoint matrix $-iF=-i\sum_jF_j$ 
and up to this order one gets
\begin{eqnarray}
\varepsilon_\pm&=&\pm\frac{1}{36}[\kappa^2(331776\kappa^6\Lambda^2+81(\Lambda^2-2)^4-\\\nonumber
&&2304\kappa^4(23\Lambda^4-72\Lambda^2)+32\kappa^2(1109\Lambda^6\\\nonumber
&&-900\Lambda^4-324\Lambda^2)]^{1/2}
\end{eqnarray}

\end{document}